\documentclass{article}
\usepackage{graphicx}
\usepackage{bm}
\usepackage{amsfonts,amsbsy}

\title{
\Large
A glimpse of fluid turbulence from the molecular scale
\footnote{
Preprint of an article published in International Journal of Modern Physics C, Vol.25, No.8 (2014) 1450034,
DOI: 10.1142/S012918311450034X.
Open Access 
\protect\\
http://www.worldscientific.com/doi/abs/10.1142/S012918311450034X (published version).
}
}

\author{
Teruhisa S. Komatsu\footnote{{\it Present Address}:Laboratory for Computational Molecular Design, RIKEN QBiC, Kobe 650-0047, Japan}
,Shigenori Matsumoto,\\
Takashi Shimada
and Nobuyasu Ito
\\
\small
Department of Applied Physics, The University of Tokyo,\\
\small
 Hongo, Bunkyo, Tokyo 113-8656, Japan.
}
\date{}

\begin{document}
\maketitle

%%%%%%%%%%%%%%%%%%
\begin{abstract}
Large scale molecular dynamics simulations of
 freely decaying turbulence in three-dimensional space are reported.
Fluid components are defined from the microscopic states
 by eliminating thermal components from the coarse-grained fields.
The energy spectrum of the fluid components is observed to scale reasonably well according to
 Kolmogorov scaling determined from the energy dissipation rate and the viscosity of the fluid,
 even though the Kolmogorov length is of the order of the molecular scale.
\end{abstract}

\section{Introduction}

Starting from appropriate constitutive equations of motion --- not from molecular scale ---
 often gives clear insight to the system.
However, such approaches also have some drawbacks:
 The underlying constitutive equations might be valid only in the limited conditions,
 and they are often violated or undetermined in the new, intermediate, or extreme conditions.
Molecular descriptions are much more robust in this regard,
 but the problem of the enormously large scale gap between microscopic
 and macroscopic (or mesoscopic) scales remains.

A calculation of molecular dynamic motion to study various equilibrium
 and nonequilibrium phenomena was pioneered in the mid twentieth century.
Encouraged by the fact that even the small $32$-particle system of Alder and Wainwright \cite{Alder}
 captures the tail of the branch in larger systems,
 researchers have developed methods of molecular dynamics (MD) simulation
 with the aid of the exponential growth of computational power, the so-called Moore's law \cite{Moore}.
For example, MD simulations of $10^4$- to $10^5$-particle systems have been employed
 to study hydrodynamics at low Reynolds number in two dimensions \cite{RapaportClementi,Rapaport,Ishiwata}
 and heat conduction in two- and three-dimensional particle systems \cite{Shimada,Murakami}.

Recent developments in parallel computers are further
 accelerating the speed of growth beyond Moore's law,
 and huge ($>$$10^{8}$-particle system) simulations
 are becoming realistic,
 and the realms reachable from the molecular scale are getting broader.
Although the molecular scale simulations are computationally intensive,
 they could become reasonable, realistic approaches
 to some extreme classes of phenomena such as nanofluidics in high Reynolds number,
 where the underlying constitutive equations for a continuum description might fail.
Because recent engineering applications are increasingly confronted with such extreme systems,
 the role of molecular scale simulations is becoming more important.

Here we focus on molecular scale simulations of turbulent flow,
 which has been one of the most challenging targets in hydrodynamics.
In the fluctuating turbulent fluid, energy currents sustain the hierarchical scale structures,
 which can usually be approached by starting from coarse-grained phenomenological descriptions
 and assuming clear separation between macroscopic (hydrodynamic) and microscopic (molecular) scales.
However, the validity of such phenomenological approaches can be doubtful,
 especially when the hydrodynamic scale is comparable to the molecular scale,
 where fluid fluctuations compete with molecular scale fluctuations.
Then a clear starting point from the molecular scale would have merit.
The aim of the present research is not to replace all fluid simulations with molecular scale simulations
 but rather to test challenging simulations and check whether molecular scale simulations are feasible.
Such simulations would play a complementary role under certain extreme conditions in which the continuum fluid description fails.

Resolving turbulent flow from the molecular scale requires 
 a huge number of molecular particles,
 and larger systems need longer simulation time scale.
The present parallel computer architecture allows us to treat larger systems,
 although it is not yet easy to treat longer time simulations.
Thus it would be a useful exercise to examine how well
 we can approach turbulence from the molecular scale by using current computer systems.

\section{Methods}
\subsection{Molecular dynamics simulation}

In this paper, we focus on MD simulation of
 freely decaying fluid flow starting from an initial velocity profile,
 the Taylor-Green vortex (TGV)\cite{TGV,Brachet},
 which is one of the most well known benchmark systems for fluid simulation.
The velocity profile of the TGV,
\begin{equation}
\bm{u}^\mathrm{TG}(x,y,z)
=
U_0
\left(
\begin{array}{c}
\sin(2\pi x/L) \cos(2\pi y/L) \cos(2\pi z/L)\\
-\cos(2\pi x/L) \sin(2\pi y/L) \cos(2\pi z/L)\\
0
\end{array}
\right),
\label{eq:TGV}
\end{equation}
 is composed of twisted vortex pairs in an $L^3$ rectangular periodic box,
 and its energy per mass is $U_0^2/8$.

In the molecular description the fluid is composed of $N$ ($\approx 10^8$) identical particles of mass $m$.
The time developments of the $i$-th particle's position $\bm{r}_i$ and momentum $\bm{p}_i$ are described as
\begin{equation}
\begin{array}{rcl}
(d/dt) \bm{r}_i &=& \bm{p}_i/m,\\
(d/dt) \bm{p}_i &=& - \sum_{j\ne i} \partial_{\bm{r}_i} V(|\bm{r}_i-\bm{r}_j|) + \bm{f}_\mathrm{wall},
\end{array}
\label{eq:eqofmotion}
\end{equation}
with simple repulsive model for the interaction potential $V(r)$,
\begin{equation}
V(r)=Y (r_c-r)^{a}
\end{equation}
for $r<r_c$ and $V(r)=0$ for $r\ge r_c$,
 where $r$ is the distance between the particles.

By utilizing the symmetry of the TGV,
 the present MD simulation is performed in an $(L/2)^3$ rectangular box with slip boundary conditions
 at the surfaces of the box, not in the $L^3$ periodic box.
This reduces the simulation costs by a factor of $\sim$8.
Although one may have concerns about the correctness of this treatment,
 we have confirmed that almost the same energy spectra are
 obtained for the periodic boundary system and the slip boundary system ($L=400$),
 at least within the order of the turnover time scale.

To realize the slip boundary conditions at the surfaces of the box,
 each particle near the $\xi (=\{x,y,z\}$) wall is (additionally) forced
 by a self-mirror image with the interaction potential $V(r)$, i.e.,
\begin{equation}
\bm{f}_\mathrm{wall} = - \partial_{\xi_i} V(2|\xi_i-\xi_\mathrm{wall}|) \hat{\bm{e}}_\xi,
\end{equation}
where $\xi_i$ ($\xi_\mathrm{wall}$) is
the $\xi$-component of the $i$-th particle (wall) position
and $\hat{\bm{e}}_\xi$ is the unit vector in the $\xi$-direction.

The time developments of the MD system are calculated by using Hamiltonian dynamics (\ref{eq:eqofmotion})
 starting from a microscopic state of particles at $t=0$
 whose macroscopic properties correspond to a velocity field $\bm{u}^\mathrm{TG}(x,y,z)$,
 a kinetic temperature $T_0$ (with a Boltzmann constant of unity),
 and a uniform number density $\bar{\rho}$.
The initial microscopic states are prepared as follows.
The particle positions $\{\bm{r}_i\}$ are located on FCC sites
whose lattice constant is adjusted to reproduce the specified number density $\bar{\rho}$.
The particle momentum $\bm{p}_i$ is taken from a Gaussian distribution with the amplitude $\sqrt{m T_0}$.
The origin of time is defined just after the execution of short transient simulations (typically $t_\mathrm{tr}=10$)
 from these configurations.
Then the initial conditions for the molecular fluid
 are prepared by adding position-dependent velocity components $m\bm{u}^\mathrm{TG}(\bm{r}_i)$ to each particle's $\bm{p}_i$.

Unless otherwise specified,
 we have used the parameter set $m=1$, $r_c=1$, $a=2$, and $Y=500$.
Under the conditions presented in this paper,
  particles with these parameters are sufficiently hard to ensure that
 the contribution of the potential energy to the total energy was less than a few percent.
The time developments of the configuration $\{\bm{r}_i(t),\bm{p}_i(t)\}$
 are calculated using a second order symplectic integrator
 with time step $\Delta t = 4\times 10^{-4}$.
Our simulation code was developed based on the flat MPI parallel code, MDACP\cite{mdacp},
 which can treat short-ranged interacting particle systems with a high efficiency of parallelization.
Main simulation runs utilized 256 nodes $\times$ 64 threads on SR16000M1
(see \ref{sec:costofsimulation} for detail).

\subsection{Coarse-grained field quantities}
From  microscopic information such as the positions and the velocities of the particles,
 we calculate those local averages to obtain the coarse-grained field quantities.
First, we grid the system into cubic cells
 so that a cell labeled $(n_x, n_y, n_z)$
 occupies a volume $n_\xi l_\mathrm{cell}  \le \xi < (n_\xi+1) l_\mathrm{cell}$ (for $\xi=\{x,y,z\}$).
Then the coarse-grained field quantities are defined at each center of the cell,
 $\bm{r}=((n_x+1/2) l_\mathrm{cell},(n_y+1/2) l_\mathrm{cell},(n_z+1/2) l_\mathrm{cell})$.
To be specific, we define
 (number) density fields
 $\rho(\bm{r}):=\sum_i w(\bm{r},\bm{r}_i) 1 /l_\mathrm{cell}^3$,
 velocity fields
 $\bm{u}(\bm{r}):=\sum_i w(\bm{r},\bm{r}_i) \bm{p}_i / m \rho(\bm{r}) l_\mathrm{cell}^3$,
 and so on,
 where the function $w(\bm{r},\bm{r}_i)$ takes the value $1$ if the particle $i$ is in the cubic cell and otherwise $0$.

\section{Results}

\begin{figure}
\begin{center}
\includegraphics[bb=0 0 1200 2400,width=6cm]{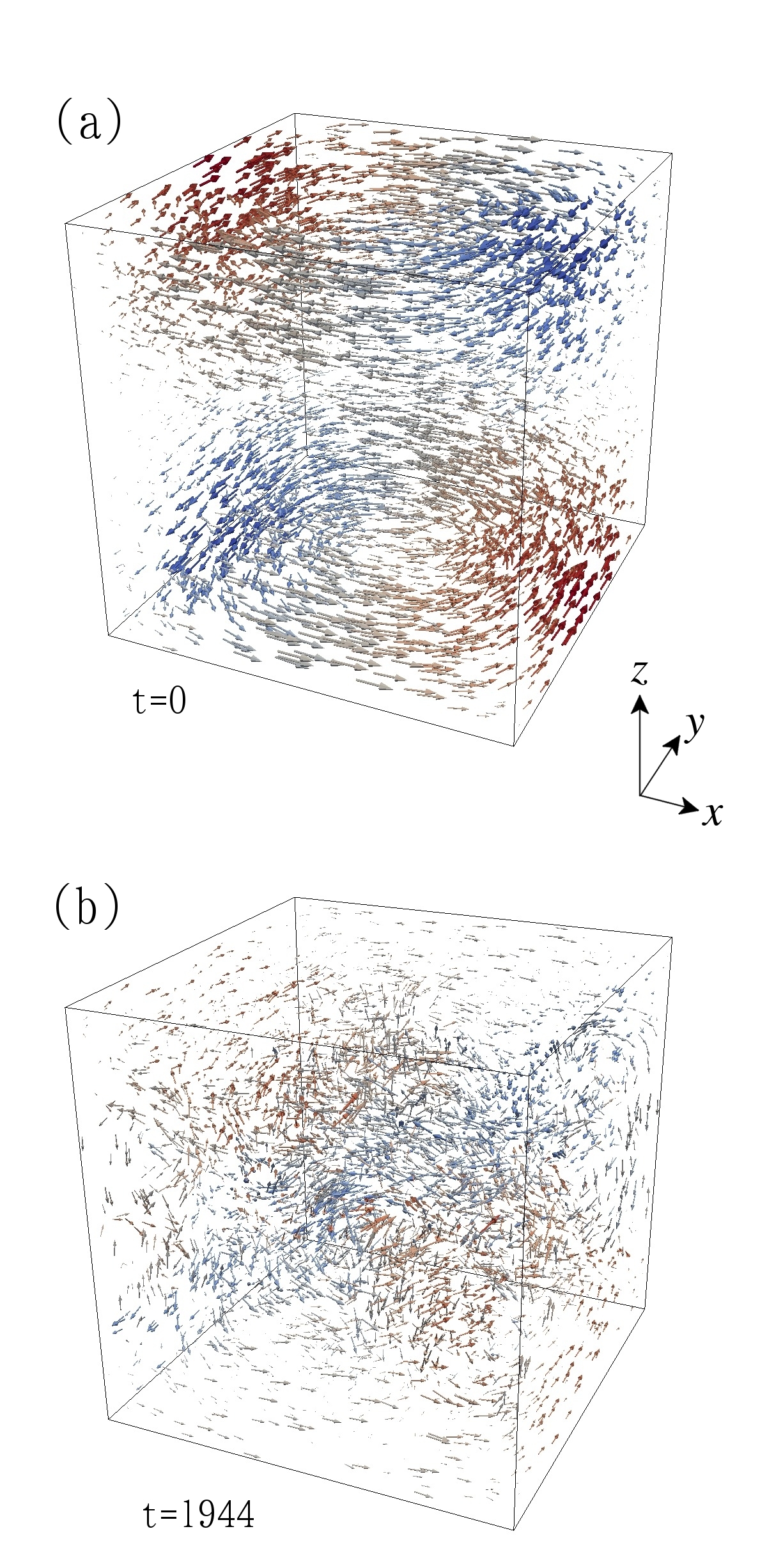}
\end{center}
\caption{
Velocity fields observed
 at (a) $t=0$ and (b) $t=1944$
 in the MD simulation box of $(L/2)^3$ ($L/2=1080,N=3.779136\times 10^8,U_0=2,T_0=0.33,l_\mathrm{cell}=10$).
Color encodes $y$ components of the velocities.
}
\label{fig:flow3d}
\end{figure}

Although the total energy of the MD system is conserved,
 coherent ``fluid'' motion decays in the course of time.
In order to observe such a fluid property from the MD data, the fluid components must be extracted.
Hence, we observe the coarse-grained field quantities defined in the previous section.
Typical snapshots of the observed velocity fields in our MD simulation are shown in Fig. \ref{fig:flow3d}.

%FIG1

These fields in $(L/2)^3$ space are converted to those in $L^3$ space before spectrum analysis.
The ``energy'' per mass described by these coarse-grained fields is
\begin{equation}
\frac{1}{{m\bar{\rho}L^3}}\int\mathrm{d}\bm{r} \frac{m\rho(\bm{r})}{2}|\bm{u}(\bm{r})|^2,
\end{equation}
which can be written as the $k$-shell averaged power spectrum
\begin{equation}
\frac{1}{2L^3}\int\mathrm{d}\bm{r} |\bm{v}(\bm{r})|^2 = \frac{1}{2}\sum_{\bm{k}} |\tilde{\bm{v}}(\bm{k})|^2
=:\int dk E(k),
\label{eq:energySpectrumDefinition}
\end{equation}
 of the vector field
\begin{equation}
\bm{v}(\bm{r}) := \bm{u}(\bm{r}) \sqrt{\rho(\bm{r})/\,\bar{\rho}\;}
=: \sum_{\bm{k}} \tilde{\bm{v}}(\bm{k}) \mathrm{e}^{i\bm{k}\cdot\bm{r}}.
\end{equation}

\begin{figure}
\begin{center}
\includegraphics{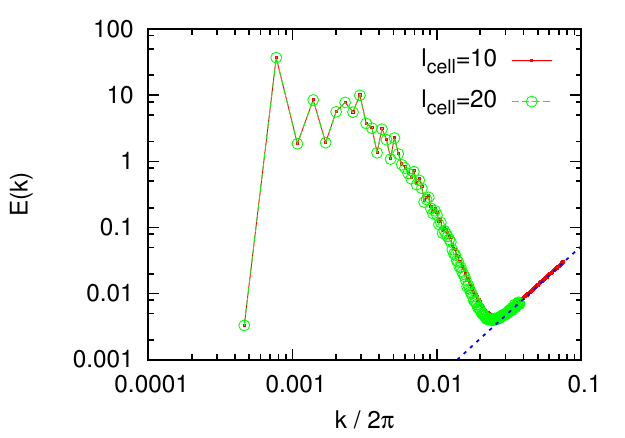}
\end{center}
\caption{
Energy spectrum for $L/2=1080$, $T_0=0.33$, $\bar{\rho}=0.30$, $U_0=2.0$, and $t=2160$.
Two cases of cell size $l_\mathrm{cell}=10$ (small red squares) and $20$ (green circles) are shown.
The broken line is $(3T/m\bar{\rho})(k/2\pi)^2$ with $T=0.52$, which is the temperature obtained
from the averaged variance of the molecular velocities in each cell ($l_\mathrm{cell}=10$).
}
\label{fig:rawSpectrum}
\end{figure}

In the energy spectrum $E(k)$ as shown in Fig. \ref{fig:rawSpectrum},
 thermally equilibrated (equipartitioned) spectrum proportional
 to $k^2$ is observed in the small-scale (high-wave-number) region
 after some transient time.
When $l_\mathrm{cell}$ is varied, only the higher-wave-number cutoff of the spectrum is varied,
 while lower-wave-number spectrum is unchanged.
This suggests that the $k^2$ branch corresponds to small-scale random thermal modes.
In the following, we take $l_\mathrm{cell}=10$ to observe the entire scale range
 from microscopic to hydrodynamic in the energy spectrum.

%FIG2

\begin{figure}
\begin{center}
\includegraphics{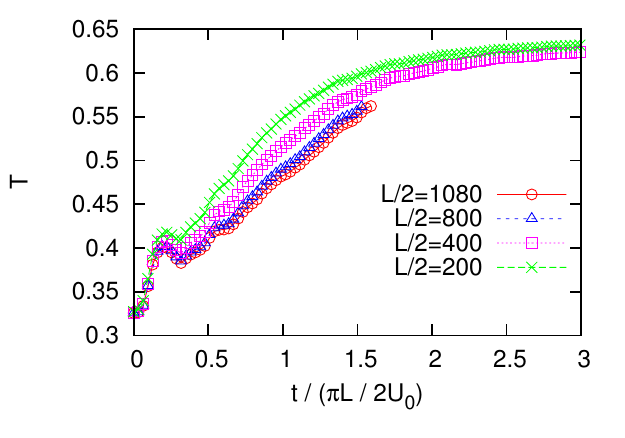}
\end{center}
\caption{
Time developments of temperature $T$ determined from the
 averaged variance of the molecular velocities in each cell.
$T_0=0.33$, $\bar{\rho}=0.30$, and $U_0=2.0$.
(Note that this definition of temperature might not be reliable during the early stage.)
At the final equilibrium state ($t\rightarrow\infty$), the temperature would become
$T_\infty=T_0+(2/3)(m/8)U_0^2=0.66$.
}
\label{fig:Temperature}
\end{figure}

When the equipartition with temperature $T$ is satisfied,
 $\tilde{\bm{v}}(\bm{k})$ becomes $\bm{k}$-independent and
 the kinetic energy for each cell (of mass $m\bar{\rho}\,l_\mathrm{cell}^3$)
is equal to $(3/2)T$, i.e.
\begin{equation}
(m\bar{\rho}\,l_\mathrm{cell}^3/2)\;
(L/l_\mathrm{cell})^3\;
|\tilde{\bm{v}}(\bm{k})|^2
=(3/2)T,
\end{equation}
where we note that $\sum_{\bm{k}} 1 = (L/l_\mathrm{cell})^3$.
Then by multiplying the density of states in the $k$-shell,
the energy spectrum in thermal equilibrium is expressed as
\begin{equation}
E_\mathrm{eq}(k) = (4\pi k^2/k_\mathrm{min}^3)\,(3T/2m\bar{\rho}L^3)
= (3T/m\bar{\rho})\,(k/2\pi)^2,
\label{eq:thermalbranch}
\end{equation}
where $k_\mathrm{min}:=2\pi/L$.
The values of temperature $T$ estimated from the variance of molecular velocities in each cell
 are consistent with those estimated from $E(k)$ (see Fig. \ref{fig:rawSpectrum}).
The time developments of $T$ are shown in Fig. \ref{fig:Temperature},
 where the molecular fluid seems to be heated  during the course of relaxation.

%FIG3

Eliminating these thermally equilibrated components, we estimate the fluid components from the MD data.
For example, the energy of the fluid components, $E^\mathrm{fluid}$, is extracted
from the fitting of the function
$
\int_0^{k} dk' E(k')
$
to the form $a k^3 + E^\mathrm{fluid}$ in the high-wave-number region.

We also observe the enstrophy, which is a measure of the vorticity field $\mathrm{rot}\bm{u}(\bm{r})$.
The enstrophy spectrum is defined by
\begin{equation}
\int\mathrm{d}k \Omega(k):=\int\mathrm{d}\bm{r} |\mathrm{rot}\bm{u}(\bm{r})|^2/(2L^3).
\end{equation}
Similarly to the energy, the enstrophy of the fluid components, $\Omega^\mathrm{fluid}$,
 is extracted from the fitting to $a' k^5 + \Omega^\mathrm{fluid}$,
where the terms $\mathrm{rot}\bm{u}(\bm{r})$ can be calculated in Fourier space.

\begin{figure}
\begin{center}
\includegraphics{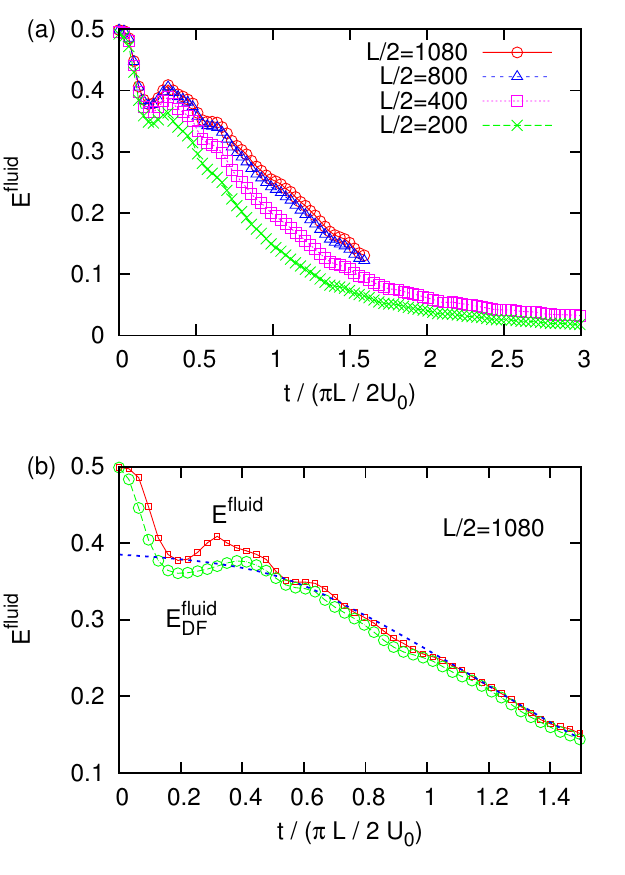}
\end{center}
\caption{
(a) Time developments of $E^\mathrm{fluid}$ for $\bar{\rho}=0.30$, $T_0=0.33$, and $U_0=2.0$.
(b) Comparison between $E^\mathrm{fluid}$ (squares) and $E^\mathrm{fluid}_\mathrm{DF}$ (circles).
The two curves coincides at later stages.
The broken line is the estimation of fluid energy
 by $E^\mathrm{fluid}(t_*)+2\nu\int_t^{t_*}\mathrm{d}t'\Omega^\mathrm{fluid}(t')$
 with $\nu=0.72$ fitted during the later stage.
}
\label{fig:Energy}
\end{figure}

The time developments of $E^\mathrm{fluid}$ for various system sizes $L$ are shown in Fig. \ref{fig:Energy}(a).
One can observe that decay of the fluid energy roughly corresponds to the increase in the temperature in Fig. \ref{fig:Temperature}.

In Fig. \ref{fig:Energy} (b),
 the divergence-free (DF) components of energy are also shown based on
 the vector decomposition
\begin{equation}
\tilde{\bm{v}}_\mathrm{RF}(\bm{k}):=\bm{k}[\bm{k}\cdot\tilde{\bm{v}}(\bm{k})]/k^2,\;
\tilde{\bm{v}}_\mathrm{DF}(\bm{k}):=\tilde{\bm{v}}(\bm{k})-\tilde{\bm{v}}_\mathrm{RF}(\bm{k}),
\end{equation}
in which $\tilde{\bm{v}}_\mathrm{RF}(\bm{k})$ are the rotational-free (RF) components and
$
|\tilde{\bm{v}}(\bm{k})|^2=
|\tilde{\bm{v}}_\mathrm{RF}(\bm{k})|^2+
|\tilde{\bm{v}}_\mathrm{DF}(\bm{k})|^2
$
.

Comparing $E^\mathrm{fluid}$ and the energy of the DF components ( $E^\mathrm{fluid}_\mathrm{DF}$) 
 in Fig.\ref{fig:Energy} (b),
 one notes that the major contribution to the energy comes from the DF components.
The bumpy shape during the early stage is mainly due to the RF components (at least for this case) and
 comes from sound wave propagation in the system.
Since the present $U_0$ is comparable to the speed of sound,
 the density profile becomes rather inhomogeneous during the early stage of the simulation,
 but during the later stage, it  relaxes toward the homogeneous one.
Thus  approximating the fluid as incompressible might be allowed during the later stage.

%FIG4

\begin{figure}
\begin{center}
\includegraphics{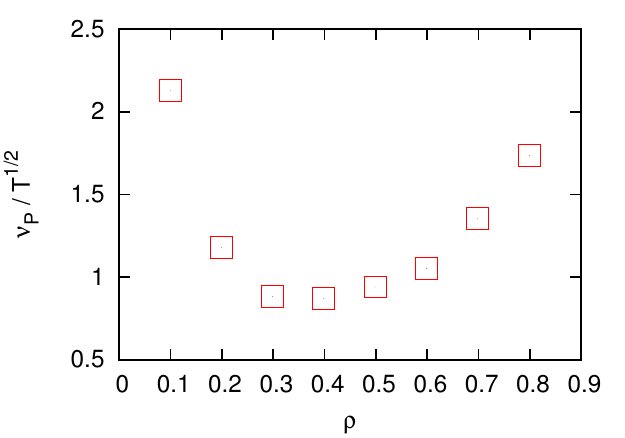}
\end{center}
\caption{
Dynamic viscosity $\nu_\mathrm{P}$ obtained from Poiseuille flow
 simulated in a periodic box ($L_\mathrm{w}\times 100\times 100$),
 where fixed wall conditions are imposed by a Langevin thermostat in the flat plate region
 ($8\times 100\times 100$).
The velocity profile (in the central region excluding the vicinity of the wall) is fitted
 to the parabolic form $u_z(x)=(g/2\nu_\mathrm{P})\;x\,(L_\mathrm{w}-x)$,
 where $g$ is the gravitational acceleration and $L_\mathrm{w}$ is the wall-to-wall length
 ($g=0.001$, $L_\mathrm{w}=50$, and $T=0.5$).
Because the interaction potential is hard for the given temperature $T$,
 the dynamic viscosity $\nu_\mathrm{P}$ would be scaled as $\sqrt{T}$.
}
\label{fig:viscosity}
\end{figure}

Since $E^\mathrm{fluid}$ has been determined, the energy dissipation rate $\epsilon$ can be estimated.
In case of an incompressible Navier-Stokes fluid,
$\epsilon$ and $\Omega^\mathrm{fluid}$ are related by
\begin{equation}
\epsilon = 2\,\nu\,\Omega^\mathrm{fluid},
\end{equation}
where $\nu$ is the dynamic viscosity.
By assuming this relation holds during the later stage for the present system,
 the values of $\nu$ can be fitted from the data.
These values of $\nu$ are consistent with those of $\nu_\mathrm{P}$ obtained
 in the other series of simulations for Poiseuille flow (see Fig.\ref{fig:viscosity}).

%FIG5

\begin{figure}
\begin{center}
\includegraphics{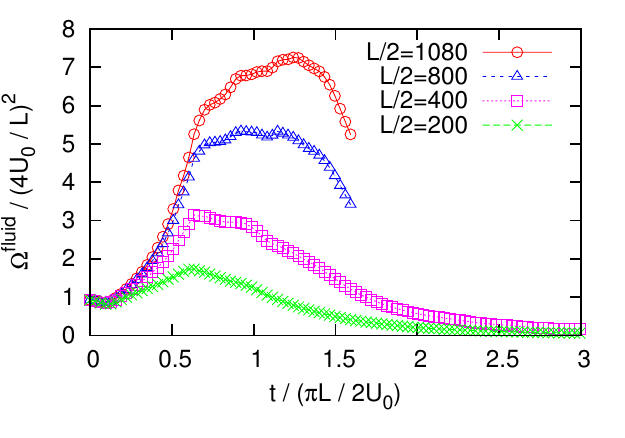}
\end{center}
\caption{
Time developments of $\Omega^\mathrm{fluid}$ for
 $\bar{\rho}=0.30$,
 $T_0=0.33$,
 and $U_0=2.0$.
}
\label{fig:Enstrophy}
\end{figure}

As shown in Fig. \ref{fig:Enstrophy}, $\Omega^\mathrm{fluid}$ grows in time
 and takes a maximum value at time $t_*$ (comparable to the order of the turnover time $\pi L/2 U_0$).
Around $t=t_*$, the production of small-scale structure in the fluid mode and the relaxation
 toward equilibrium balance in some sense.
The maximum values of enstrophy normalized by its initial one
 increase with $L$, which suggests that larger systems can produce wider scale structures.

%FIG6

\begin{figure}
\begin{center}
\includegraphics{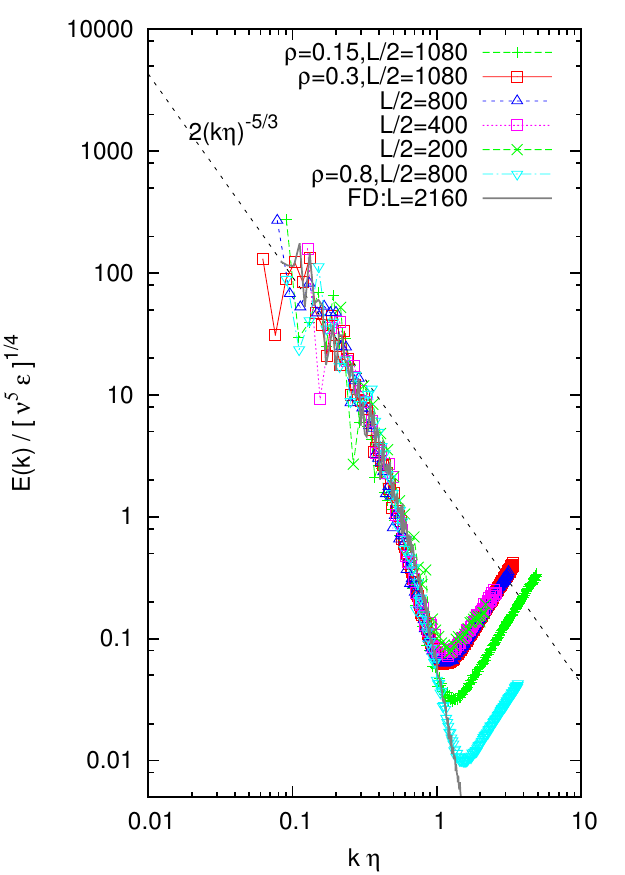}
\end{center}
\caption{
Scaled energy spectrum
$E(k)/(\nu^5\epsilon)^{1/4}$.
The parameters are shown in Table \ref{table}.
The dotted line $2 (k\eta)^{-5/3}$ is drawn to guide to eye.
The FD result is obtained from the simulation of the incompressible Navier-Stokes equation
 ($L=2160$, $\nu=0.7$, and $U_0=1.7$) by using a pseudo spectrum method ($200^3$ modes)
 with a simple 3/2 padding aliasing free method \protect\cite{dealiasing}.
}
\label{fig:Kspectrum}
\end{figure}

Now let us study the energy spectrum at $t=t_*$,
 the time at which the flow is expected to be most turbulent.
Figure \ref{fig:Kspectrum} shows the energy spectra for the various $L$ values listed in Table \ref{table}
(where, for clarity, only the higher-wave-number range is shown),
 in which the spectrum obtained from continuum fluid dynamics (FD) simulation is also shown.
First, one notes that the spectra scale reasonably well (including the FD results) according to the
 Kolmogorov scaling determined by $\epsilon$ and $\nu$ \cite{K41,Batchelor,KSpectrumExp}.
Second, in the present case, crossover scales to thermal modes are rather close to
 the Kolmogorov length
\begin{equation}
\eta := (\nu^3/\epsilon)^{1/4},
\end{equation}
 and clear deviation of the (dissipation) spectrum is not noticed.
Although the system size of the present simulation is still limited,
 $E(k)$ seems to approach Kolmogorov spectrum $(k\eta)^{-5/3}$ around $k\eta\leq 0.2$.
Additional large-scale simulations
 would demonstrate the spectrum more clearly.

%FIG7

%TABLE1
\begin{table}[ht]
\caption{
Parameters and observables at $t=(\pi L/2 U_0) \tilde{t}_*$.
Listed are fluid energy $E$ and enstrophy $\Omega$ determined from the spectra
 (see $E^\mathrm{fluid}$ and $\Omega^\mathrm{fluid}$ in the text),
energy dissipation rate $\epsilon$ from the decay of $E$,
temperature $T$ from the variance of velocities in each cell,
dynamic viscosity $\nu:=\epsilon/2\Omega$,
Kolmogorov length $\eta:=(\nu^3/\epsilon)^{1/4}$,
and Reynolds number $R:=(L/2) U/\nu$
 with  characteristic velocity $U:=(2E/3)^{1/2}$.
For all MD simulations in this table, $U_0=2$ and $T_0=0.33$.
}
\begin{tabular}{|c|c|c|c|c|c|c|c|c|c|c|c|c|c|c|c|c|c|c|c|c|c|c|c|c}
\hline
$L/2$
 & $\bar{\rho}$
 & $\tilde{t}_*$
 & $E$
 & $10^{4}\Omega$
 & $10^{4}\epsilon$
 & $T$
 & $\nu$
 & $\eta$
 & $R$
\\
\hline
\hline
$1080$
 & $0.15$
 & $0.83$
 & $0.25$
 & $0.51$%\times 10^{-4}$
 & $1.09$%\times 10^{-4}$
 & $0.490$
 & $1.08$
 & $10.4$
 & $410$
\\
\hline
$1080$
 & $0.30$
 & $1.24$
 & $0.20$
 & $1.00$%\times 10^{-4}$
 & $1.44$%\times 10^{-4}$
 & $0.515$
 & $0.72$
 & $7.1$
 & $548$
\\
\hline
$800$
 & $0.30$
 & $0.92$
 & $0.26$
 & $1.33$%\times 10^{-4}$
 & $1.91$%\times 10^{-4}$
 & $0.480$
 & $0.72$
 & $6.6$
 & $464$
\\
\hline
$400$
 & $0.30$
 & $0.64$
 & $0.31$
 & $3.15$%\times 10^{-4}$
 & $4.61$%\times 10^{-4}$
 & $0.447$
 & $0.73$
 & $5.4$
 & $249$
\\
\hline
$200$
 & $0.30$
 & $0.64$
 & $0.26$
 & $6.93$%\times 10^{-4}$
 & $10.84$%\times 10^{-4}$
 & $0.477$
 & $0.78$
 & $4.6$
 & $106$
\\
\hline
$800$
 & $0.80$
 & $0.95$
 & $0.25$
 & $1.53$%\times 10^{-4}$
 & $3.18$%\times 10^{-4}$
 & $0.435$
 & $1.04$
 & $7.7$
 & $314$
\\
\hline
\end{tabular}
\label{table}
\end{table}

\section{Discussion}

In this paper, we have reported MD simulations of Taylor-Green vortex
 to explore fluid turbulence from a simple molecular starting point.
We have determined the energy and enstrophy of the fluid mode from the cell-averaged MD data
 by eliminating thermal modes associated with microscopic molecular motion.
The extracted observables of the fluid mode during the later stage imply that the fluid can be treated as  incompressible.
The obtained energy spectrum scales well according to Kolmogorov scaling,
 even though the spectrum around the Kolmogorov length ( $k\eta\sim 2\pi$ )
 is dominated by the thermal modes.
Below around $k\eta\leq 0.2$, the spectrum seems to approach a power law, which
 might indicate a glimpse of turbulence (or that we have reached  the smallest unit of turbulence)
 from the molecular scale.

For  efficient usage of computational power to resolve turbulent flow,
 a rather strong (supersonic) velocity field is initially imposed;
 this contributes to narrowing the separation between molecular  and fluid scales.
The resulting Kolmogorov length is of the order of several particle diameters,
 which could be of the order of the microscopic scale, the mean free path of the particles.
As seen in Table \ref{table}, $\eta$ increase with $L$ as $\eta\propto L^{1/4}$.
This relationship holds  because the dissipation rate is mainly determined by
 the macroscopic time scale $L/U_0$, and thus roughly $\epsilon \propto L^{-1}$.

Usually, the Kolmogorov length is sufficiently large compared to the microscopic scale,
 as one can simply assume the separation of scales between  fluid  and  microscopic ones.
For example, from a rough estimation
 for atmospheric gas ($\nu\sim 10^{-2}$ m$^2$/s and $m\bar{\rho}\sim 1$ kg/m$^3$)
 under  conditions such that
 the large-scale velocity,  large-scale length, and  energy dissipation rate
 are
 $U\sim 10$ m/s, $L\sim 10$ m, and $\epsilon\sim U^3/L$
 (solely determined from the large-scale motion),
respectively, the resulting Kolmogorov length is $\eta\sim 10^{-2}$ m,
 which is $10^5$ times the order of the mean free path.
Correspondingly, the height of the scaled energy spectrum for the thermal branch
 at $k\eta=1$ is estimated to be \textless$10^{-15}$
 from Eq.(\ref{eq:thermalbranch}) divided by $(\nu^5\epsilon)^{1/4}$.
Thus, it is rather hard to observe the thermal branch
 at the tail of the dissipation spectrum under usual conditions.

However, in this study,
 the Kolmogorov length becomes comparable to the microscopic scale,
 i.e., turbulent fluctuations compete with thermal fluctuations,
 but the energy spectrum still seems to be merely a superposition of or crossover between fluid and thermal modes.
This observation is suggestive for nanofluidics in  extreme conditions.
Although apparent interference between molecular  and fluid scales has not been noticed
 in the present observation,
 further detailed studies may yield some insight.
An ingenious setup to precisely observe microscopic states
 in macroscopically steady turbulent states will be essential to proceed.

It is a benefit of MD simulation that we can observe  turbulence directly from the molecular scale.
Observations from the molecular scale up to and beyond the Kolmogorov length
 will supplement observations based on continuum descriptions
 and be expected to become a complementary method, especially for systems under special or extreme conditions.

In addition to those academic interests,
 direct simulation from the molecular scale will have merits in some engineering situations,
 for example, the flow of complex fluids that contains phase transitions,
 impact ruptures, or coalescence.
For such complex flows, ambiguities may  remain in the constitutive equations and so
 clear starting points from the molecular scale will attract much attention.

%%%%%%%%%%%%%%%% Acknowledgment %%%%%%%%%%%%%%%%
\section*{Acknowledgments}

We wish to thank M. Miyama and Y. Murase for helpful advices.
Numerical simulations were partly carried out by the use of the Plasma Simulator
 at the National Institute for Fusion Science and supported by the NIFS Collaboration Research programs (NIFS10KNSS014).
This work was partly supported by Award No. KUK-I1-005-04 made by King Abdullah University of Science and Technology (KAUST).

%%%%%%%%%%%%%%%% Appendix %%%%%%%%%%%%%%%%%%%%
\appendix

\section{Costs of simulation}
\label{sec:costofsimulation}

The simulation time for $L/2=1080$ system upto $t=2700$
 using 256nodes $\times$ 64 threads on SR16000M1(POWER7) was 12 hours.
Simulation at early stage was inefficient
 because of unbalanced loads on each thread due to the density inhomogeneity.
File I/O cost is rather expensive, e.g. it took 900 sec (55msec per thread) to write down single snapshot
 of particle configuration (17GB binary file),
 and it seems to grow with the system size.
Thus efficient algorithm and hardware of file I/O (if particle-resolution is necessary)
 should be implemented for the simulation using huge number of threads.

When the system size $L$ is increased,
 number of particles increases with $L^3$
 and the characteristic time scale roughly increases with $L$ for constant $U_0$.
The former $O(L^3)$ can be tackled by using huge scale multi-core system.
The latter $O(L)$ has direct influence on the turn-around time of single simulation.

%%%%%%%%%%%%%%%% References %%%%%%%%%%%%%%%%%%%%

\end{document}